\documentclass[prl,aps,twocolumn,superscriptaddress]{revtex4}
\usepackage{graphicx,amssymb,amsmath,color,psfrag}
\emergencystretch=\maxdimen
\usepackage{amsthm}
\usepackage{amsfonts}
\usepackage{algorithmic}
\usepackage{enumerate}
\usepackage{latexsym}
\usepackage{amsmath}
\usepackage{amssymb}
\usepackage[dvipdfmx,colorlinks,linkcolor=blue,anchorcolor=blue,citecolor=blue,urlcolor=blue]{hyperref}

\begin{document}

\title{Triplet $p+ip$ pairing correlations in the doped Kane-Mele-Hubbard model: A quantum Monte Carlo study}
\author{Tianxing Ma}
\affiliation{Department of Physics, Beijing Normal University,
Beijing 100875, China\\}

\affiliation{Beijing Computational Science Research Center,
Beijing 100084, China}

\author{Hai-Qing Lin}
\affiliation{Beijing Computational Science Research Center,
Beijing 100084, China}

\author{J. E. Gubernatis}
\affiliation{Theoretical Division, Los Alamos National Laboratory,
Los Alamos, NM 87545 USA}

\date{\today}

\begin{abstract}
{\ By using the constrained-phase quantum Monte Carlo method, we performed a systematic study of the pairing correlations in the ground state of the doped Kane-Mele-Hubbard model on a honeycomb lattice.
We find that pairing correlations with $d+id$ symmetry dominate close to half filling, but pairing correlations with $p+ip$
symmetry dominate as hole doping moves the system below three-quarters filling. We correlate these behaviors of the pairing correlations with the topology of the Fermi surfaces of the non-interacting problem. 
We also find that the effective pairing correlation is enhanced greatly as the interaction increases, and these superconducting correlations are robust against varying the spin-orbit coupling  strength.
Our numerical results suggest a possible way to realize spin triplet superconductivity in doped honeycomb-like materials or ultracold atoms in optical traps.
}
%with extended-$s$ symmetry. However, as the system
%size or the on-site Coulomb interaction increases, the long-range part
%of the $d+id$ pairing correlation decreases and tends to vanish in the
%thermodynamic limit. An inclusion of nearest-neighbor interaction $V$,
%either repulsive or attractive, has a small effect on the extended-$s$
%pairing correlation, but strongly suppresses the $d+id$ pairing correlation.  }
\end{abstract}
\maketitle

%71.10.Fd Lattice Fermion models (Hubbard model, etc.)
%74.20.Mn Nonconventional mechanisms
%74.20.Rp Pairing symmetries (other than s-wave)

\section{ Introduction}
%Correlated materials have remarkable properties and transitions between distinct, competing phases with dramatically different electronic and magnetic orders. Among these transitions and properties, what is happening in the vicinity of various magnetic orders and superconductivity is one of the most actively asked but not yet completely answered questions about heavy-Fermion materials\cite{Wolfle2011}, organic superconductors \cite{Kagawa2005}, and such high temperature superconductors as doped cuprates\cite{Moon2010}  and iron-pnictides\cite{Johnston2010}.
Recently, a new %electron physics
 research field  has emerged in condensed matter physics which is based on the findings that a spin-orbit interaction can lead to topological electronic phase transitions  \cite{Kane2005} and the electron-electron interactions can produce these transitions \cite{Sun2009,Liu2010,Raghu2008,Wen2010,Rachel2010,Yu2011,Herbut2013,HHH2013}. This new field makes it natural to ask what remarkable new properties and transitions might occur between distinct, competing correlated topological electron phases when the spin-orbit interaction is present. Various researchers have proposed \cite{Raghu2008,Wen2010,Rachel2010,Yu2011,Herbut2013,HHH2013} the Hubbard model with spin-orbit interactions (the Kane-Mele-Hubbard model) on a honeycomb-like lattice, as a starting point for answering these questions.
%To date, the study of electron correlation effects in this model has mainly followed two routes: (1) investigating the effects on the intriguing non-interacting properties of honeycomb  material such as graphene\cite{Novoselov2005}, silicene\cite{Vogt2012}, and germanene\cite{Cahangirov2009} and (2) investigating the effects on the topological physics of the non-interacting electrons in the presence of a spin-orbit coupling. When the electron-electron interactions are not zero, more recent studies suggest that superconductivity appears when the model is doped away from half-filling. In particular, the experiments of Sebastian et al.\cite{Sebastian2010} argue that near a quantum critical point(QCP), doping (or applying a pressure) in the vicinity of a quantum phase transition enhances the electron-electron interaction which in turn enhances the superconducting transition temperature.

Indeed the density of states of the non-interacting version of the model on a honeycomb lattice has
special features that point to special properties in its interacting version. % of the model.
In particular, if the hopping is nearest neighbor,
a van Hove singularity (VHS) exists in the non-interacting model's density of states (DOS) at $3/4$ electron filling \cite{Ma2010}.
This feature prompted several groups \cite{Wen2011,Nandkishore2012,Thomale2012}
to suggest singlet superconductivity in the interacting model,
and several others \cite{Schaffer2007,Jiang2008,Baskaran2010,Ma2011,Zhong2014} to suggest $d+id$  pairing in the low doped case.
Numerical simulations point to other novel states of matter tied to the strength of the electron-electron interaction and spin-orbit coupling.
In particular, if $U$ is the strength of the local Coulomb interaction and $t$ the hopping amplitude to nearest neighbor sites, the Hubbard model on a honeycomb lattice is known to have a semimetal phase at small $U$ and an antiferromagnetic one at large $U$ \cite{Paiva2005},
and ground-state quantum Monte Carlo (QMC) simulations performed at half-filling suggest a
transition from a semimetal to an insulating antiferromagnetic state insulator % when  $U/t$ is small, and
when $U/t$$\approx$3.5\cite{Herbut2013,Paiva2005}. %, regardless of the present of spin orbit coupling.
%Just before this transition, a possible gapped spin liquid state was proposed \cite{Meng2010,Hohenadler2011}
%in {\color{red} both the Hubbard model on a honeycomb lattice \cite{Meng2010} and the Kane-mele-Hubbard model\cite{Hohenadler2011}},
%and then strongly questioned \cite{Sorell a2012}.
On the other hand,  other quantum Monte Carlo studies, such as the finite temperature determinant quantum Monte Carlo method  \cite{Ma2010}, find that strong antiferromagnetic fluctuations dominate
around half filling and strong ferromagnetic correlations dominate at less than 3/4-filling.
Accordingly, the Kane-Mele-Hubbard model on a honeycomb lattice is an opportunity to study a variety of novel
quantum critical and thermal fluctuations of itinerant charge carriers, with doping and interacting strengths
being very relevant control parameters that may lead to new topologically-constrained states of matter.%\cite{Muramatsu2014}.

In this paper we study quantum fluctuations in the Kane-Mele-Hubbard model via ground state quantum Monte Carlo simulations. Because of the already observed strong magnetic fluctuations, we ask whether superconducting fluctuations exist in novel forms. A common view, supported by recent experiments \cite{Tacon2011}, is that the magnetic fluctuations provide a glue for pairing and generally lead to unconventional superconductivity \cite{Scalapino2012}. Relative to what we learned from past numerical studies on this model \cite{Ma2014}, at least two important questions remain: What happens to the pairing for dopings below the van Hove singularity? and How does the spin-orbit coupling affect the pairing at any doping? In conjunction with these two questions, we can also ask whether $p+ip$ spin triplet superconductivity is possible. A topological superconductor with $p+ip$
pairing symmetry is the most elemental way in which a non-Abelian topological state can emerge as the
ground state of a many-body system \cite{Nayak2010} and thus provides an ideal platform to construct a possible quantum computer \cite{Stern2013}.

%{\color{blue} Topological quantum computation continues to be one of the most exciting approaches to construct
%Topological quantum computation exploits topological phases of matter whose quasiparticle excitations obey non-Abelian braiding statistics.
%
%. Though great progress has been achieved, this novel state of matter has not yet been realized in any real material\cite{Sarma2006}}.

We address the questions stated above by using the constrained-phase quantum Monte Carlo method \cite{Ortiz1997,Schmidt1999,Schmidt2001,Schmidt2003,Sarsa2003,Zhang2003}. This method allows us to perform simulations not possible with more commonly used quantum Monte Carlo methods. As we show, our results reveal rich physics when both the spin-orbit coupling and the electron-electron interactions are strong. Between half-filling and the filling at the VHS, we find that the $d+id$ pairing dominates. Below the lower band VHS filling, we find strong triplet $p+ip$ superconducting correlations. Further, the effective pairing correlation is enhanced greatly as the interaction increases, and these superconducting correlations are robust against varying the spin-orbit coupling. On the basis of these results, we suggest that triplet $p+ip$ topological superconductivity might be realizable in doped honeycomb-like materials \cite{Novoselov2005,Vogt2012,Cahangirov2009} or ultracold atoms in optical traps \cite{Gurarie2005,Wu2007,Tarruell2012}.

%{\it Model and Methods}:
\section{ Model and Methods}
The Kane-Mele-Hubbard model \cite{Kane2005} on the honeycomb lattice is
\begin{equation}
\label{KMH-model}
H=-t \sum_{\langle {\bf ij} \rangle} c^{\dag}_{{\bf i}\sigma}c_{{\bf j}\sigma}^{\phantom{\dag}} + i\lambda_{SO}\sum_{\langle\langle {\bf i,j} \rangle\rangle,\sigma}\sigma\nu^{\phantom{\dag}}_{{\bf ij}}c^{\dag}_{{\bf i}\sigma}c_{{\bf j}\sigma}^{\phantom{\dag}}+ U\sum_{{\bf i}} n_{{\bf i}\uparrow}^{\phantom{\dag}}n_{{\bf i}\downarrow}^{\phantom{\dag}}
\end{equation}
($c_{{\bf i}\sigma}^{\phantom{\dag}}$)
where $t$, $U$, and $\lambda_{\rm{SO}}$ are the nearest neighbor (NN) hopping energy, the strength of the on-site repulsion, and the next-nearest neighbor (NNN) spin-orbit coupling strength, respectively.
Here $c_{{\bf i}\sigma}^{\phantom{\dag}}$($c_{{\bf i}\sigma}^{\dag}$) annihilates (creates) an electron with spin $\sigma$ on site ${\bf i}$ and $\nu_{\bf ij}=\pm 1$ depending on if the
electron makes as ``right" or ``left" turn when going from site ${\bf i}$ to site ${\bf j}$ on the honeycomb lattice. We assume periodic boundary conditions. %We compute various ground-state properties of this model by performing simulations with the constrained-phased quantum MOnte Calro method. %\cite{Kane:prl05,Kane_2:prl05}

The constrained-phase quantum Monte Carlo method (PCPMC) \cite{Ortiz1997,Schmidt1999,Schmidt2001,Schmidt2003,Sarsa2003,Zhang2003} is a generalization of the constrained-path method (CPMC) \cite{Zhang1995,Zhang1997,Carlson1999} and is an analog of the fixed-phase generalization of the fixed-node diffusion Monte Carlo method \cite{Ortiz1993}. The PCPMC method has yielded very accurate results for the ground state energy and other ground state observables for various strongly-correlated lattice models \cite{Chang2008,Chang2010a,Chang2010b,Xu2011} and for atoms, molecules, and nuclei \cite{Schmidt1999,Schmidt2001,Schmidt2003,Sarsa2003}. The PCPMC method is sometimes called the phaseless method \cite{Zhang2003}.

Briefly, for a Hamiltonian $H=H_0+H_1$, the constrained-path method, as most ground state quantum Monte Carlo methods, projects to the ground state $|\psi_0\rangle$ from some initial state $|\psi_T\rangle$ via
\begin{align}\label{Projection}
|\psi_0 \rangle =& \lim_{\tau\rightarrow\infty} e^{-\tau H}|\psi_t\rangle \approx e^{-\Delta\tau H_0}
e^{-\Delta\tau H_1} \times \nonumber \\
&\quad e^{-\Delta\tau H_0} e^{-\Delta\tau H_1}\cdots
e^{-\Delta\tau H_0}e^{-\Delta\tau H_1}|\psi_T\rangle
\end{align}
where $H_0$ is the non-interacting part of the Hamiltonian and  $H_1$ is the interacting part. Here, $H_0$ is the hopping and spin-orbit parts of \eqref{KMH-model} and $H_1$ it is the Hubbard $U$-term. In the constrained-path method, a Hubbard-Stratonovich transformation is applied to each exponential of the interaction. With the consequence of introducing distributions of auxiliary scalar fields into the problem, converting the exponential of the interaction into an exponential of a non-interacting term,  depends on imaginary-time dependent external fields. %For a Hubbard interaction these fields couple to the $z$-component of the spin at the lattice sites.
With an initial state being an ensemble of Slater determinant, the constrained-path method propagates one ensemble of Slater determinants into another. The individual propagating determinants are called random walkers.
% and each walker represents a different configuration of the Hubbard-Stratonovich fields.
%This propagating determinant is called a random walker,
%The simulation is initialized with a population of such wlakers.
%This propagating determinant becomes a random walker in the space of Slater determinants.
%and a large population of such walkers $\{ |\phi_i\rangle \}$  are propagated from the same initial state.
%The different configurations associate a weight $w_i$ with each walker. The initial state is also used to importance sample the Hubbard-Stratonovich fields.
At any step in the projection, the projected state is a weighted sum of the walkers, $|\psi\rangle \approx \sum_i w_i |\phi_i\rangle$. %Our initial state was the solution to the non-interacting problem.

After a sufficient number of steps, the method begins to sample walker weights that represent those of the ground state wave function. For most Fermion simulations these weights are generally not all non-negative and hence do not represent a probability distribution amenable to Monte Carlo sampling. This is called the sign problem.

The constrained-path method approximately handles the sign problem, which is caused by a broken symmetry in the space of Slater determinants, by eliminating any random walker as soon as $\langle\phi_i|\psi_T\rangle < 0$. The presence of the spin-orbit interaction in the Hamiltonian means the ground state cannot be real. To ensure samples come from a real, non-negative distribution, the constrained-phase approximation generalizes the constrained-path condition: With a phase $\theta$ defined by
\[
e^{i\theta} \equiv \langle\phi|\psi_T \rangle / |\langle\phi|\psi_T\rangle|
\]
two simple forms of constrained-phase method follow \cite{Ortiz1997} from replacing the walker by
$
|\phi\rangle \leftarrow \cos(\theta) e^{-i\theta}|\phi_i\rangle
$
and eliminating the walker if $\Re{\langle\phi_i|\psi_T\rangle} < 0$ or by
$
|\phi\rangle \leftarrow e^{-i\theta}|\phi_i\rangle
$
which makes $\langle\phi_i|\psi_T\rangle > 0$.
%We found the variances of our expectation values were a bit smaller with the first choice.

\begin{widetext}
\begin{center}
\begin{table}[h]
\caption{  Comparisons of the energies (total $E_T$, kinetic $E_k$, and potential $E_p$), double occupancies and correlation
functions ( $S$ and $S_d$ the spin and charge density structure factors ) of the two-dimensional $U=4.0$ Hubbard
model on different lattices.
Different rows correspond to ED, CPMC, and PCPMC results. The multi-column heading of $4\times 4$ lattice with $ 5\uparrow\,5\downarrow$ refers to the Hubbard model on a square lattice, that of $4\times 4$ lattice with $ 4\uparrow\,4\downarrow$ to the Hubbard model on a square lattice in a magnetic field with 1/4 of flux quanta per plaquette, and that of $2\times 3^2$ KMH model with $ 9\uparrow\,9\downarrow$ to the Kane-Mele Hubbard model with $\lambda=0.1$.}
\begin{tabular}{c|c|c|c||c|c|c|c||c|c}
\hline\hline
    & \multicolumn{3}{c||}{ $4\times 4$ lattice with $ 5\uparrow\,5\downarrow$ } & \multicolumn{4}{c|}{ $4\times 4$ lattice with $ 4\uparrow\,4\downarrow$ in a magnetic field } & \multicolumn{2}{c}{$2\times 3^2$ KMH model with $ 9\uparrow\,9\downarrow$}\\ \cline{2-10}
     & $E_T$ &  S$(\pi,\pi)$   & S$_{d}(\pi,\pi)$  & $E_T$  & $E_k$ & $E_p$  & Double occupancy & $E_T$ & Double occupancy    \\
\hline
ED      & -19.581    & 0.73      & 0.506   &   -19.783 &  -21.715  & 1.932 & 0.0030185 &  -14.51   &  0.143 \\
CPMC    & -19.580(1) & 0.730(1)  & 0.507(1) &    &    &  &  & & \\
PCPMC   & -19.580(4) & 0.731(1)  & 0.508(1)  &    -19.779(1) & -21.712(2)    & 1.933(1)  & 0.0030188 &   -14.35(1) &0.148(2)\\
\hline\hline
\end{tabular}
\end{table}
\end{center}
\end{widetext}

Here we used the first constraint. Other than this difference in constraint, and the algorithmic details of the constrained-phase method are the same as those in the constrained-path method \cite{Zhang1995,Zhang1997,Carlson1999}. When no phase problem is present, the constrained-phase method in fact reduces to the constrained-path method. When no sign problem is present, the constrained-path method is exact.

In Table~I we show a comparison of the CPMC and PCPMC methods with exact diagonalization (ED) method for several different lattices (square and honeycomb), electron dopings, and spin orbit interaction strengths. Both the CPMC and PCPMC methods agree with each other and with the ED results for energies, double occupancies, and spin-spin correlations. The key point is that the PCPMC allows accurate simulations, as for two cases in Table~I, not possible with CPMC or other fixed-node-like methods.

\begin{figure}[b]
\includegraphics[scale=0.45]{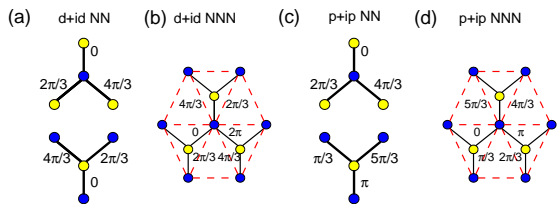}
\caption{(Color online) Phases of the pairing symmetries on the honeycomb
lattice: (a) $d+id$ with NN (b) $d+id$ with NNN, (c) $p+ip$ with NN and (d) $p+ip$ with NNN. Here different colored dots denote sites of different sub-lattices A and B.} \label{Fig:Pairing}
\end{figure}

%Because the Hubbard-Stratonovich transformation transforms the interacting problem into a non-interacting one, expectation values values of many-body correlation functions, for example, products of two creation operators with two destruction others, are easily evaluated with the use of Wick's Theorem which reduces these expectation values to sums of products of equal-time, single-particle Green's functions. This Green's function is a natural product of the method.

%Guided by the predictions in \cite{XXX,YYY,ZZZ},
%Guided by the RPA
We computed superconducting correlation functions for four different pairing symmetries: (a) the $d+id$ NN, (b) the $d+id$ NNN %\textbf{JG: added an N. OK?},
(c) the $p+ip$ NN, (d) and the $p+ip$ NNN symmetries, whose form factors are illustrated in Fig.~\ref{Fig:Pairing}. These different pairing symmetries are distinguished by different phase shifts upon 60$^o$ or 120$^o$ rotations. We now define the vectors $\delta_{{\ell}}^{\phantom{\dag}}$ and $\delta_{{\ell}}^{\prime}$ to denote the NN  and NNN  inter-sublattice connections where ${\ell}=1,2,3$ or ${\ell}=1,2,3,4,5,6$
label the different directions.
On either sublattice, the NNN-bond $p+ip$ and $d+id$ wave pairings have the following form factors,
\begin{align}
%\text{$ES$-wave} &\text{:}&\ f_{ES}(\delta_{l})=1,~l=1,2,3 \\
\ f_{p+ip}(\delta^\prime_\ell)&=e^{i(\ell-1)\frac{ \pi}{3}}, \nonumber \\
  f_{d+id}(\delta^\prime_\ell)&=e^{i(\ell-1)\frac{2\pi}{3}},~\ell=1,2,...6,
\end{align}
while the singlet NN-bond $d+id$ pairing has the form factor
\begin{equation}
%\text{$ES$-wave} &\text{:}&\ f_{ES}(\delta_{l})=1,~l=1,2,3 \\
\ f_{d+id}(\delta_\ell)=e^{i(\ell-1)\frac{2\pi}{3}},~{\ell}=1,2,3.
\end{equation}
The NN-bond $f_{p+ip}$ is different for an A and B sublattice: for sublattice A,
\begin{equation}
\ f_{p+ip}(\delta_\ell)=e^{i(\ell-1)\frac{2\pi}{3}},~ \ell=1,2,3,
\end{equation}
while for sublattice B, there is a $\pi$ phase shift (Fig. \ref{Fig:Pairing}c)
\begin{equation}
\ f_{p+ip}(\delta_\ell)=e^{i(\ell-1)\frac{2\pi}{3}+i\pi},~\ell=1,2,3 .
\end{equation}
%{\color{blue}
 The $p+ip$ NN and the $p+ip$ NNN defined above have the same symmetry but different forms in $k$-space and real space. Both are topological nontrivial spin triplet superconducting  states. As their form factors indicate,
 their real (imaginary) parts are symmetric about the x-axis (y-axis) but asymmetric about the y-axis (x-axis). Thus they have a $p_x+ip_y$ symmetry, which is simply called $p+ip$.%}

The functional form of the pairing correlation function we computed is
\begin{eqnarray}
C_{\alpha }({\bf{r=R_{i}-R_{j}}})=\langle \Delta _{\alpha }^{\dagger }
({\bf i})\Delta _{\alpha }^{\phantom{\dagger}}({\bf j})\rangle ,
\end{eqnarray}
where $\alpha$ stands for the pairing symmetry and the corresponding order
parameter $\Delta_{\alpha }^{\dagger }({\bf i})$\ is
\begin{eqnarray}
\Delta_{\alpha }^{\dagger }({\bf i})\ =\sum_{\ell}{\color{blue} f_{\alpha}^{*}}
(\delta_{{\ell}})(c_{{{\bf i}}\uparrow }c_{{{\bf i}+\delta_{{\ell}}}\downarrow }-
c_{{{\bf i}}\downarrow}c_{{{\bf i}+\delta_{{\ell}}}\uparrow })^{\dagger},
\label{Delta}
\end{eqnarray}
%In this last equation,  $f_{\alpha}(\delta_\ell)$ is the form factor.

\begin{figure}[tbp]
\includegraphics[scale=0.45]{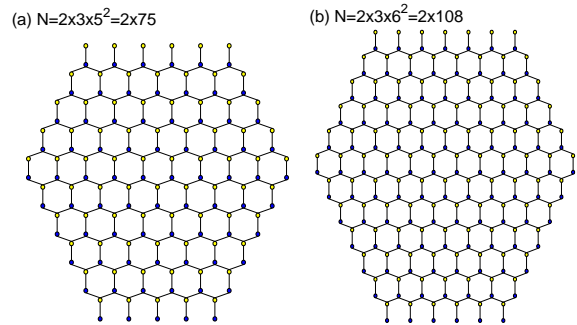}
\caption{The lattice geometries for the $2\times75$ ($2\times3\times 5^2$) and $2\times108$ ($2\times3\times 6^2$) honeycomb lattices.}
\label{Fig:Figstructure}
\end{figure}

%{\it Results and discussion}:
\section{ Results and discussion}

\begin{figure}[tbp]
\includegraphics[scale=0.4]{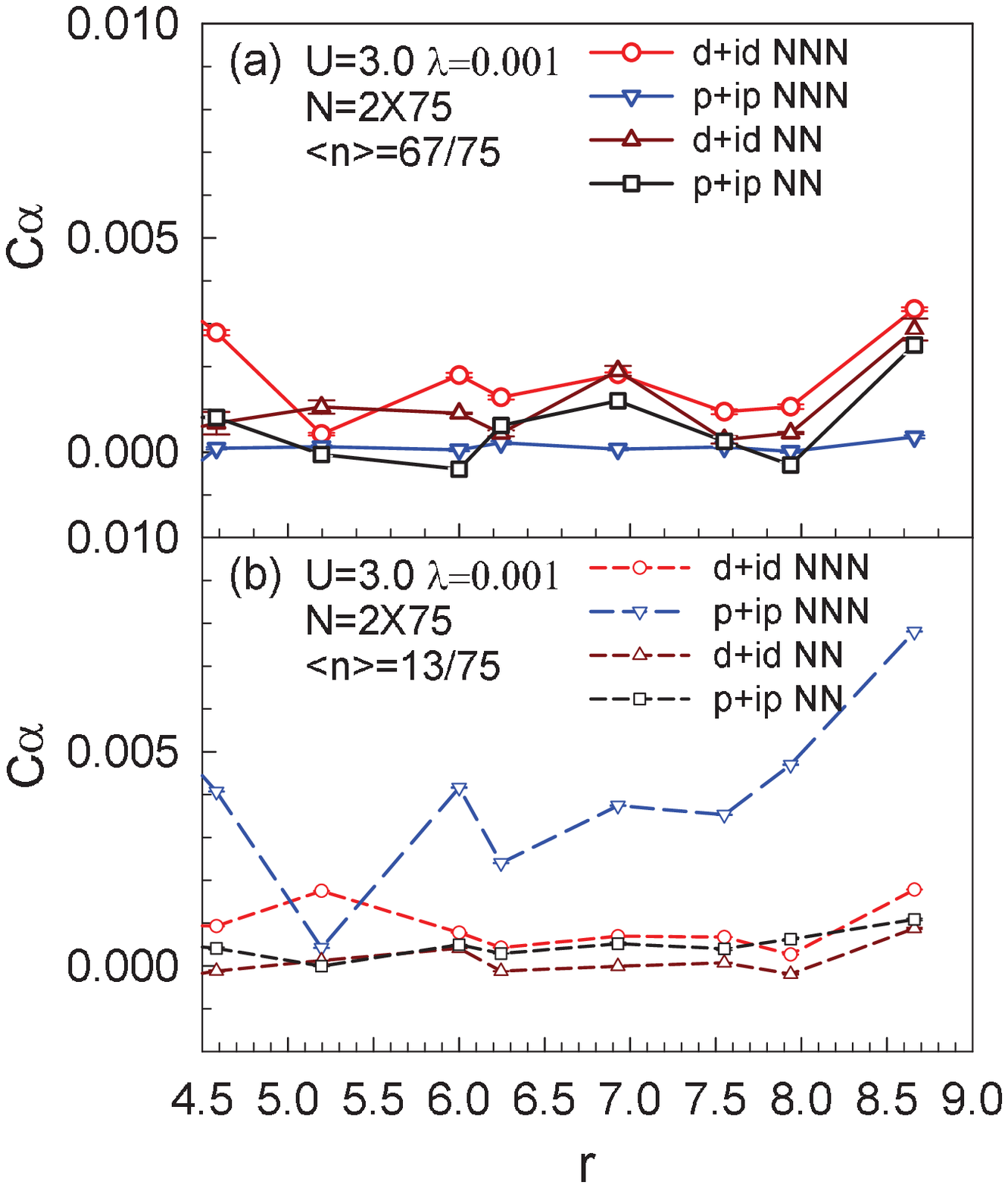}
\caption{(Color online) Pairing correlation $C_{\alpha}$ as a
function of distance $r=|{\bf{R_{i}-R_{j}}}|$ for different pairing symmetries on
the double-75 lattice with $\langle n\rangle=67/75$ (a) and
with $\langle n\rangle=13/75$ (b) at $U=3.0$ and $\lambda=0.001$.} \label{Fig:SOC001}
\end{figure}

\begin{figure}[b]
\includegraphics[scale=0.4]{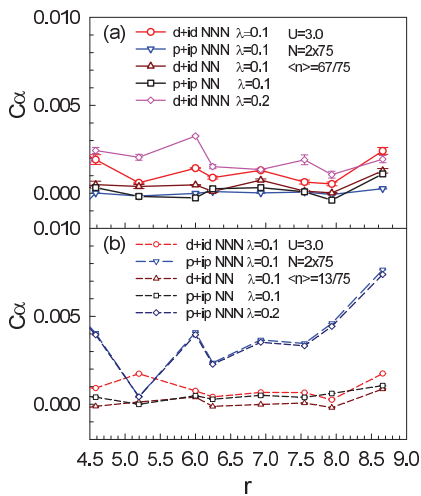}
\caption{(Color online) Pairing correlation $C_{\alpha}$ as a
function of distance $r=|{\bf{R_{i}-R_{j}}}|$ for different pairing symmetries on
the double-75 lattice with $\langle n\rangle=67/75$ (a) and
with $\langle n\rangle=13/75$ (b) at $U=3.0$ and $\lambda=0.1$.} \label{Fig:SOC01}
\end{figure}

%that is, two hopping coupled triangle lattices each with 75 sites.
%The honeycomb can be described as two interpenetrating triangular sublattices%. Hence the total number of sites in our $2 \times 75$ lattice is 150.
Our numerical simulations are mainly performed on a $2 \times 75$ lattice and a $2 \times 108$ lattice, which is shown in Fig. \ref{Fig:Figstructure}.
The honeycomb can be described as two interpenetrating triangular sublattices, which are marked by different colors, blue and yellow, in Fig. \ref{Fig:Figstructure}.
 The lattices shown preserve most geometric symmetries of the triangular Bravais lattice, and are often adopted in research on graphene.
In Fig.~\ref{Fig:SOC001}, we compare the long-range part of pairing
correlations for the $d+id$ and $p+ip$ pairing symmetries on
double-75 lattices at $U=3.0$ and $\lambda=0.001$. The simulations were
performed for the closed-shell cases corresponding to fillings of (a) $\langle n\rangle$=$67/75$  and  (b) $\langle n\rangle$=$13/75$.  For finite-size non-interacting problems, these fillings correspond to non-degenerate ground states. Simulations performed at these fillings exhibit much smaller statisical variance in the measurements than those performed at other fillings.
%This lattice setting preserves most geometric symmetries of the triangle lattice.

The first is a filling above that of the VHS; the other is below. As previous
results in doped graphene without spin orbit coupling \cite{Jiang2008,Ma2011}, above the VHS $C_{d+id}(r)$ is larger than $C_{p+ip}(r)$
for most long-range distances between electron pairs. At $\langle n\rangle$=$13/75$, while below, the NNN-bond $C_{p+ip}(r)$ tends to be larger than the others. Here $r$=$|{\bf{R_{i}-R_{j}}}|$. For the present lattice, $r=9$ corresponds to a separation of lattice sites about half the cell size.
From another work \cite{Ma2010}, we know that ferromagnetic fluctuations dominate electron fillings below the VHS, and antiferromagnetic correlations dominate around half filling.
Thus, the $p+ip$ superconducting pairing appears favored by ferromagnetic correlations.

\begin{figure}[t]
\includegraphics[scale=0.4]{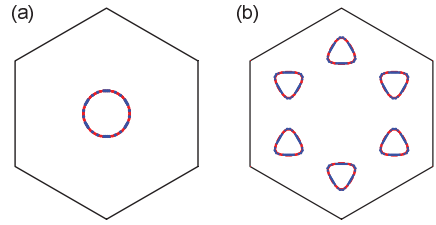}
\caption{(Color online) The Fermi surface at $\langle n\rangle=13/75$ (a) and $\langle n\rangle=67/75$ (b) with $\lambda=0.001$ (red color) and $\lambda=0.1$ ( blue color).
} \label{Fig:Fermi}
\end{figure}

%We note in Fig.~\ref{Fig:SOC001}b a difference between the present model and study of pairing correlations in the Hubbard model \cite{Zhang1995,Zhang1997}. In the Hubbard model, near half filling, the Coulomb interaction was observed to suppress the pairing correlations. Here, presumably because low doping corresponds to small double occupancy, this interaction has little negative effect on the pairing.  Because of the Trotter approximation \eqref{Projection} in our simulation method the range of Coulomb interaction strength is restricted to the weak and intermediate coupling regimes.

We also show results for larger spin-orbit couplings in Fig.\ref{Fig:SOC01}. % that are an other of magnitude larger those those in Fig.~\ref{Fig:SOC001}.
For the larger spin-orbit coupling of $\lambda=0.1$, the $d+id$ correlations continue to dominate over other pairing forms at $\langle n\rangle=67/75$, and the
$p+ip$ correlations continue to dominate over other pairings at $\langle n\rangle=13/75$. One interesting point is that when compared to those in Fig.~\ref{Fig:SOC001}, the pairing correlations with $d+id$ symmetry are enhanced as the spin-orbit coupling increases,
while the $p+ip$ pairing correlations are suppressed, but only slightly, by increasing the spin-orbit coupling. Thus the correlations we are finding are quite robust, at least with respect to the spin-orbit interaction.

 The observed dependencies of the pairing correlations on filling and spin-orbit coupling can be understood from the Fermi surface topologies. The Fermi surface with different spin-orbit couplings are shown in Fig.~\ref{Fig:Fermi} at two electron fillings: (a) $\langle n\rangle=13/75$  and (b)$\langle n\rangle=67/75$. In each sub-figure the results for $\lambda=0.001$ are indicated by a red contour and those for $\lambda=0.1$ by a blue contour.  At the low filling, $\langle n\rangle=13/75$ in Fig.~\ref{Fig:Fermi}(a), the Fermi surface is a small circle around the $\Gamma$ point, which favors the $p+ip$ pairing. At the higher filling in in Fig.~\ref{Fig:Fermi}(b), $\langle n\rangle=67/75$, the nesting vector of Fermi surface
is near to the antiferromagnetic vector, which favors singlet $d+id$ pairing.  In both sub-figures, the Fermi surface varies only very slightly as $\lambda$ increases. The behaviors of the Fermi surfaces thus correlate with the behaviors of pairing correlations in Figs.~\ref{Fig:SOC001} and \ref{Fig:SOC01}, where the pairing correlations an insensitivity to the scale of spin-orbit coupling.

\begin{figure}[b]
\includegraphics[scale=0.4]{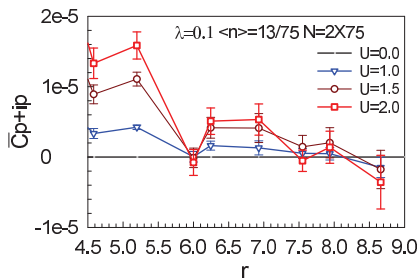}
\caption{(Color online) The vertex contribution $\bar{C}_{p+ip}$ as a
function of distance $r$ for different $U$ on
the double-75 lattice with $\langle n\rangle=13/75$ and $\lambda=0.1$.} \label{Fig:FigU}
\end{figure}

In order to examine the intrinsic pairing interaction in our finite system, we extracted the vertex contribution ${\bar C}_\alpha$ from $C_{\alpha}$. Doing so requires subtracting  uncorrelated single-particle contributions, such as  $\langle c_{{i}\downarrow }^{\dag
}c^{\phantom{\dag}}_{{j}\downarrow }\rangle \langle c_{i+\delta_{l}\uparrow }^{\dag }
c^{\phantom{\dag}}_{j+\delta_{l'}\uparrow }\rangle $, from correlated many-body terms, such as $\langle
c_{{i}\downarrow }^{\dag }c_{{j}\downarrow }c_{i+\delta_{l}\uparrow}^{\dag}
c^{\phantom{\dag}}_{j+\delta_{l'}\uparrow}\rangle $. In Fig. \ref{Fig:FigU}, the vertex contribution $\bar{C}_{p+ip}$ as a
function of distance $r$ for different $U$ on the double-75 lattice with $\langle n\rangle=13/75$ and $\lambda=0.1$ is shown.
At $U=0$, the vertex contribution is zero, and at $U=1.0$, the $\bar{C}_{p+ip}$ tends to be a finite positive value over most of the long range part.
Clearly, it is interesting to see that the $\bar{C}_{p+ip}$ is enhanced greatly as the interaction increases from $U=1.0$ to $U=2.0$, over most of the long range part{\color{blue} \cite{vertex}}.
Such behavior of the vertex contribution suggests effective attractions generated between electrons and the instability toward $p+ip$ superconducting pairing in the system.
The enhanced pairing strength with the enhancement of the electron correlations indicating the interaction plays a key role in inducing superconductivity. We note that the observed behavior as a function of $U$ contrasts that observed for the two-dimensional Hubbard model \cite{Zhang1995,Zhang1997}.
%In the numerical results, the ratio of the statistical error to the pairing correlation $C_{\alpha}$ is no more than 0.5 percent,
%and most of the error bars are almost within the symbols.
%The ratio of the statistical error to the vertex contribution $\overline{V_{\alpha}}$ is no more than 3 percent. %, which can be seen in Fig.~\ref{Fig:Symmetry1} (b).
%This remark applies to all the following figures.
%Clearly in Fig. \ref{Fig:FigU}, the intrinsic pairing interaction $\bar{C}_{p+ip}$ at $U=2.0$ is a finite value over most of the long range part.
%For a comparison, the $\bar{C}_{p+ip}$ at $U=0$ and $U=1$ are also shown. At $U=0$, the  It is interesting to see that the intrinsic pairing interaction $\bar{C}_{p+ip}$ is enhanced greatly as the interaction increases.

 %while vertex contributions of $d_{xy}$, $s_{x^2+y^2}$ and $d_{x^2-y^2}$ are simply fluctuating around zero.
%{\color{blue}

\begin{figure}[h]
\includegraphics[scale=0.45]{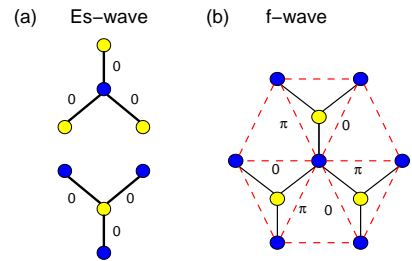}
\caption{(Color online) Phases of the pairing symmetries on the honeycomb
lattice (a) $Es$-wave and (b) $f$-Wave. Here different colored dots denote different sub-lattices.} \label{Fig:Esf}
\end{figure}

\begin{figure}[b]
\includegraphics[scale=0.4]{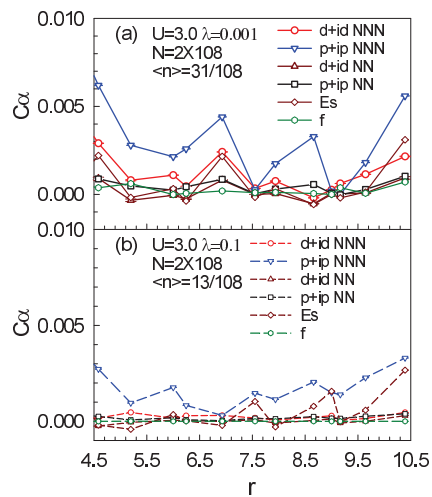}
\caption{(Color online) Pairing correlation $C_{\alpha}$ as a
function of distance $r=|{\bf{R_{i}-R_{j}}}|$ for different pairing symmetries on
the double-108 lattice with (a) $\langle n\rangle=19/108$  and
with (b) $\langle n\rangle=13/108$  at $U=3.0$ and $\lambda=0.1$.} \label{Fig:N108}
\end{figure}

To establish more firmly the robust presence of the % topological
triplet $p+ip$ pairing correlations, we calculated the superconducting correlations for two more pairing symmetries, extended s-wave (Es-wave) and $f$-wave, at low electron fillings for a larger lattice size.
The phases of the pairing forms of $Es$-wave and $f$-wave are shown in Fig.~\ref{Fig:Esf}(a) and (b). From Fig.~\ref{Fig:Esf},
\begin{eqnarray}
f_{Es}(\delta_{\ell})=1,~{\ell}=1,2,3;
\end{eqnarray}
and
\begin{eqnarray}
f_{f}(\delta'_{\ell})=\frac{[1+(-1)^{\ell}]}{2}\pi,~{\ell}=1,2,3,4,5,6.
\end{eqnarray}

Figure~\ref{Fig:N108} shows the various pairing correlations as a function of long-range part of the distance on a larger 2$\times$108 lattice for the closed shell fillings of $\langle n\rangle = 19/108$ (a) and $\langle n\rangle = 13/108$ with $U=3.0$ and $\lambda=0.1$. It is clear that the $p+ip$ pairing correlations are larger than the other five pairings of different symmetries and this ranking does not change as the strength of the spin-orbit coupling changes. In the Hubbard model, we note that increasing the lattice size suppresses the magnitude of the pairing correlations \cite{Zhang1995,Zhang1997}. We also note that in the Hubbard model near half filling it is the antiferromagnetic fluctuations that dominate.

%{\color{blue}
A numerical solution for the full phase diagram for the Kane-Mele-Hubbard model is computationally challenging, especially as a function of doping. Over the range of dopings and lattice sizes studied, we do however find strong triplet $p+ip$ superconducting correlations, and this pairing is robust against varying the spin-orbit coupling. Additionally, the effective pairing correlation is enhanced greatly as the interaction increases. Because of these results, we argue that robust spin triplet $p+ip$ superconductivity might be present in doped honeycomb-like materials such as doped graphene, silicene, and germane or ultracold atoms in optical traps.%}

%{\it Acknowledgment:}
\section{ Acknowledgment}
We thank F. Yang, H. Yao and A. Muramatsu for stimulating discussions.
T. Ma thanks CAEP for partial financial support. This work is supported in part
by NSCFs (Grant. Nos. 11374034, 11334012 and U1530401) and the Fundamental Research Funds for the Central Universities (Grant No. 2014KJJCB26). The work of JEG was supported by the US Department of Energy.

%\bibliography{FeAspc2,qmc,myownpaper}

\begin{thebibliography}{99}
\bibitem{vertex} We do not report results for $U>2.0$.  For $U>2.0$, we found that our estimated error for this subtraction was comparable to our estimate of the average vertex contribution. Computing the vertex contribution requires subtracting two small noisy numbers. Even for $U=2.0$, our estimated error is around sixteen to forty percent.
\end{thebibliography}

\end{document}